\documentstyle[12pt]{article}
\newcommand{\beq}{\begin{equation}}
\newcommand{\eeq}{\end{equation}}
\newcommand{\bea}{\begin{eqnarray}}
\newcommand{\eea}{\end{eqnarray}}

\begin{document}

\begin{titlepage}

\title{{\bf INTERMITTENCY  FOR   COHERENT \\
       AND  INCOHERENT  CURRENT  ENSEMBLE  MODEL}
       \thanks{ Work partly supported by the KBN grant: 2 PO 3B 08308}
               }
\author{B.Ziaja\\
        Institute of  Physics, Jagellonian University\\
        Reymonta 4, 30-059 Cracow, Poland\\
        e-mail: $beataz@ztc386a.if.uj.edu.pl$ }
\date{April 1995}
\maketitle

{\abstract
We investigate the origin of intermittency for multiparticle distribution
in momentum space, following the idea that there is a kind of power law
distribution of the space-time region of hadron emission. Using the formalism
of current ensamble model to describe boson sources we discuss intermittency
exponents for the coherent and incoherent ( chaotic)  particle  production
scheme.}
\vspace{1.0cm}

\begin{tabbing}
PACS numbers:$13.85.$Ni
\end{tabbing}

\end{titlepage}


\addtolength{\baselineskip}{0.20\baselineskip}
\section{Introduction}

Recently, several experiments \cite{l1},\cite{l2},\cite{l3} found that the
phenomenon of intermittency \cite{l4},\cite{l5} is dominated by very short
range
correlations between momenta of identical hadrons. As it is well known, the
HBT
 correlations \cite{l6} reflect the size and shape of the space-time region
 from
 which the observed identical particles are emitted \cite{l7}. Remembering
 that
intermittency is equivalent to a power law dependence of hadronic correlation
functions (~ multiplicity distributions ), one may conclude that the power law
dependence  must also be present in the distribution of space-time shapes and
sizes of the region of hadron emission \cite{l8}. The first analysis of this
problem
was done in Ref.\ \cite{l9} using the HBT formalism for purely incoherent
hadron source. The considered description of hadron emission was simplified :
totally incoherent production from the space region was assumed, and therefore
the single particle distribution could not be adequately described.
Nevertheless,
intermittency exponents obtained in this model linearly increased with  the
rank
of multiparticle distributions.

In order to correct description of the single particle distribution
the pion source should not only have a power-law density profile but also some
space-time correlations between particle emitting points. In this paper we
would
like to consider the model for HBT correlations worked out by Gyulassy,
Kauffmann and Wilson in Ref.\ \cite{l10}, known as the covariant
current formalism or the current ensemble model. It has been often used to
describe
many different aspects
of multiparticle production ( see eg.\  Ref.\ \cite{l12}, \cite{l13}, for
review
Ref.\ \cite{l7} and references therein~). Covariant current formalism is a
generalization of the HBT formalism formulated in the language of
field theory. Pion emitting regions are described as "currents" with
different phases, and incorporated into the pion field equation
in the form of current ensemble. Each current is localized around some point
called a collision site, and
the collision sites are distributed in space-time with some probability
density.
Pion emission depends on current phases: if phases
are identical, particles are emitted coherently, if they change randomly,
the pion production will be incoherent.

Current ensemble model suits our purposes very well. On the one hand, it is
a natural continuation of the treatment presented in
\cite{l9}. The quantum-mechanical source density is replaced by the power-law
distribution of collision sites. The case of coherent and incoherent
emission can be described easily.
On the other hand, current ensemble
formalism operates already on pions ( hadrons). This fact removes one serious
difficulty
which plagues the interpretation of intermittency in terms of parton
distribution.
The problem is that, even if one explains the intermittent behaviour of the
parton system, the experiment observes hadrons, not partons. Consequently,
one should take into account the hadronization process. Usually, one invokes
at this point the principle of parton-hadron duality, the phenomenon which is
in fact not understood. However, if intermittency is caused by the HBT
phenomenon,
there is no problem of hadronization and the decay of resonances: the HBT
effect
works already on hadron level, reflecting the structure of the source density.
And so does it in the current ensemble model.

Alternative approaches to the calculation of Bose-Einstein correlations
(~ for review see \cite{l9} and references therein ) incorporate also the
reaction dynamics. Hence the current ensemble model can be considered as
a plausible " first approximation" concentrated only on the effects of source
geometry on the shape of correlation functions.

In our previous studies of the current ensemble formalism \cite{l11} we have
already
proved that assuming purely coherent production of particles,  multiparticle
distributions are influenced by the scaling space-time structure of bosonic
source, and give observed intermittent results if the source number is large
enough.

    In the present paper we analyze the case of incoherent ( chaotic)
    particle production.
We apply the method of  random phase averaging
described in Ref.\ \cite{l7} and \cite{l10} to generate incoherence.
This formalism  is presented in  section 2.  In section 3 we
explain briefly how to measure intermittency parameters. In section 4 we
incorporate the scaling dependence of the production region into the model.
At first  we discuss the simple power law  distribution of collision sites
following \cite{l9}
and prove that it does give observable intermittent result.  Later
we show that the obtained results can be generalized for other source
distributions showing
the scaling behaviour. At the end of this section we describe the special
case
of the fixed source number for correlation functions.

     The aim of our discussion is to prove that in a well working physical
     model
of boson production like the current ensemble model, multiparticle
distributions
in momentum space can be strongly influenced by the scaling behaviour of the
space-time structure of bosonic source, and can show intermittent behaviour.


\section{Classical current formalism}

\subsection{Pion fields}

In this section we would like to recall the main ideas of current ensemble
model \cite{l10}. To obtain the final pion state produced by a classical
current source one should solve the following field equation for the scalar
pion field $\Phi (x)$ with the source current operator $J(x)$ ~:
\beq
(\partial_{\mu} \partial^{\mu}+m_{\pi}^{2})\Phi(x)=J(x)
\eeq
In principle, the source current $J(x)$ is an  operator coupled to the pion
fields,
and treating it as a complex space-time function is only an approximation.
Because we will not specify the conditions of production process it is
difficult to define when this approximation could be used. We assume simply
that
we are allowed in our model to replace the pion current operator $J(x)$  by its
expectation value \cite{l10}.

Solution of (1) gives the coherent final multipion state~:
\beq
\mid\Phi> = e^{-\frac{n}{2}} exp(i\int d^{3}k J({\bf k}) a^{+}( {\bf k} ))
\mid 0>,\, <\Phi\mid\Phi>=1
\eeq
where $a({\bf k})$, $a^{+}({\bf k})$ are creation and annihilation  operators,
n is the
average pion multiplicity and  $J(k)$ is the on-mass-shell Fourier transform
of $J(x)$ \cite{l7}, \cite{l10}. Pion density matrix $\rho_{\pi}$ constructed
from (2) is~:
\beq
\rho=\mid \Phi><\Phi \mid
\eeq
Multipion-inclusive distributions $P_{m}(k_{1},\ldots,k_{m})$  are then
defined by the formula~:
\beq
P_{m}({\bf k_{1},\ldots,k_{m} })=\frac{1}{\sigma_{\pi}} \frac{d^{3m}
\sigma(\pi \ldots\pi)}{d^{3} {\bf k_{1}} \ldots d^{3}{\bf k_{m}} }=
Tr[\rho_{\pi} a^{+}({\bf k_{1}}) \ldots a^{+}({\bf k_{m}})
a({\bf k_{m}}) \ldots a({\bf k_{1}})]
\eeq
\subsection{Pion source as current ensemble}

Now  we are ready to consider  the following description of pion emission:
pions are produced in space-time centers $x_{1},\ldots,x_{N}$, distributed
with some probability density $\rho(x)$. These centers can be for example an
effect
of N separate collisions "producing" pions \cite{l10}.
In this picture, the total pion source $J(x)$ would thus be a sum of N
different
currents $J_{i}(x)$~:
\beq
J(x)= \sum_{i=1}^{N} J_{i}(x) e^{i\varphi_{i}}
\eeq
where we allowed the possibility that the individual currents have different
phases. It reduces easily to the purely coherent case described in \cite{l11}
assuming that~:
\beq
\varphi_{1}= \ldots =\varphi_{N} \equiv \varphi
\eeq
We shall consider  the situation  when the $J_{i}(x)$ depend only on distance
from the individual collision site $x_{i}$. It means that, if a collision
centered in  $x = 0$ is parametrized  by $j_{\pi}(x)$ we get~:
\beq
J(x)=\sum_{i=1}^{N} j_{\pi}(x-x_{i}) e^{i\varphi_{i}}
\eeq
and the on-mass-shell Fourier transform is given by~:
\beq
J({\bf k})= j_{\pi}({\bf k}) \sum_{i=1}^{N} e^{i\varphi_{i}}
exp(i\omega_{k}t_{i}-i{\bf kx_{i}}),
\,\omega_{k}=\sqrt{ {\bf k^{2}}+m_{\pi}^{2} }
\eeq

\subsection{Multipion distributions for current ensemble. The coherent and
incoherent case.}

The density matrix $\rho_{\pi}$ averaged over the number of sources N and
positions $x_{i}$ of the pion sources is ~:
\beq
\rho_{\pi}=\sum_{N} P(N) \int d^{4}x_{1} \ldots d^{4}x_{N} \rho(x_{1})
\ldots \rho(x_{N}) \mid \Phi><\Phi\mid
\eeq
where $P(N)$ denotes the probability to find exactly N pion sources, and
$\rho(x)$
is the probability density to find the source placed at the point x.
The density $\rho(x)$
is normalized to 1. Using the formulae (2), (4) and (9) one can obtain the
inclusive multipion distribution $P_{m}({\bf k_{1}},\ldots,{\bf k_{m}})$
in the form~:
\beq
P_{m}({\bf k_{1}, \ldots,k_{m}})=\sum_{N} P(N)\int d^{4}x_{1}
\ldots d^{4}x_{N}
\rho(x_{1})\ldots\rho(x_{N}) \mid J({\bf k_{1} })\mid^{2} \ldots
\mid J({\bf k_{m} })\mid ^{2}
\eeq
Substituting $J({\bf k})$ in (10) in the form (8) one obtains~:
\bea
P_{m}({\bf k_{1}, \ldots,k_{m}})=\mid j_{\pi}({\bf k_{1} })\mid^{2} \ldots
\mid j_{\pi}({\bf k_{m} })\mid ^{2}
\sum_{N} P(N)\int d^{4}x_{1}\ldots d^{4}x_{N}\rho(x_{1})\ldots\rho(x_{N})
\nonumber\\
\sum_{i_{1}=1}^{N} \ldots \sum_{i_{2m}=1}^{N}
e^{ik_{1}(x_{i_{1}}-x_{i_{2}})}e^{ik_{2}(x_{i_{3}}-x_{i_{4}})}
\ldots e^{ik_{m}(x_{i_{2m-1}}-x_{i_{2m}})}\\
e^{i(\varphi_{i_{1}}-\varphi_{i_{2}})}e^{i(\varphi_{i_{3}}-\varphi_{i_{4}})}
\ldots e^{i(\varphi_{i_{2m-1}}-\varphi_{i_{2m}})}\nonumber
\eea
One should remember that in the scalar product of 4-dimensional vectors
$k_{i}$
and $x_{j}$ the first component of $k_{j}$  is equal to $\omega_{k_{j}}$
as a result of the on-mass-shell Fourier transform taken in (8). For current
phases
satisfying the relation  (6) we recover multiplicity distributions for the
coherent particle production from \cite{l11}. If the sources add incoherently,
we should also average (11) over the set of phases. This method of incoherence
generation has led to the name "chaotic"production  because we assume that
the current phases $\varphi_{i}$ fluctuate randomly from 0 to $2\pi$. After
phase
averaging we obtain the  formulae for the single- and double pion
distribution in the
form~:
\beq
P_{1}({\bf k_{1}})=\mid j_{\pi}({\bf k_{1} })\mid^{2} <N>
\eeq
\beq
P_{2}({\bf k_{1},k_{2}})=\mid j_{\pi}({\bf k_{1} })\mid^{2} \mid j_{\pi}
({\bf k_{2} })\mid^{2}
(<N^{2}>\\
+<N(N-1)>\mid\rho({\bf k_{1}-k_{2} })\mid^{2})
\eeq
One should notice that for $N = 1$ one gets the coherent field results
Ref.\
\cite{l11}. The produced particles are then uncorrelated. Hence we exclude the
case $N = 1$  from our further investigations.


\section{Intermittency exponents}

Intermittency is equivalent to the power law dependence of hadronic
moments in momentum space. The hadronic moments we would like to consider
in this paper are factorial moments obtained by integration of
multiparticle distributions and cumulants obtained by integration of
correlation functions in the finite region of  size $\delta$.
Generally, one studies
the behaviour of the factorial moments which tests the whole n-particle
distribution function. However, the cumulants have the advantage of testing
the genuine n-particle correlations, and so it is always interesting to
investigate
their contribution to the higher order correlations (see e.g. \cite{l14}).

Scaling dependence of the moments is described by the intermittency exponents.
After integration of pion multiplicities
$P_{m}({\bf k_{1}},\ldots,{\bf k_{m}})$ in the finite region
of size $\delta$, one obtains a kind of series in $\delta$~:
\beq
F_{m}(\delta)=\int_{0}^{\delta} P_{m}({\bf k_{1}, \ldots,k_{m}}) d^{3}k_{1}
\ldots d^{3}k_{m}
= \sum_{j} a_{j}(L\delta)^{-\alpha_{j}}
\eeq
where $\alpha_{j}$ are intermittency indices for the factorial moment
$F_{m}$,
$a_{j}$ are the weights and the
length L is introduced for dimensional reasons. In experiment we observe
the
term dominating in (14).

    {\sl So, for our purposes we define the intermittency exponent $f_{m}$
    to be
equal to $\alpha_{j}$ taken from the term dominating in (14).} We should
notice
it need not to be the term with max/min($\alpha_{j}$) because of the
 weights
$a_{j}$.
In the similar way one can get the cumulant intermittency exponent
$\nu_{m}$.

There is not much experimental data concerning the higher order moments.
The existing ones show intermittency exponents increasing with  rank of
multiparticle distributions \cite{l15}. The NA22 group measured also
the third order cumulant with the preliminary result $\nu_{3}=2\nu_{2}$
\cite{l14}.


\section{Scaling behaviour of the distribution $\rho(x)$}

\subsection{Scaling distribution of production centers}

In this section we show that the scaling distribution of collision sites
$\rho(x)$ (~ defined in (9)~ ) can  produce the scaling of momentum
distributions
$P_{m}$.
We consider the simplest example of the distribution $\rho(x)$ with the
assumption
that all particles were produced at the same moment $t_{0}$~:
\beq
\rho(x)=\delta ( t-t_{0} )\rho_{S}(\mid {\bf x} \mid)
\eeq
and the space part follows a power law~:
\beq
\rho_{S}({\bf x})=L^{-\alpha} \mid{\bf x}\mid^{\alpha-3}, 0<\alpha<2,
\alpha \neq 1
\eeq
where the length L was introduced for dimensional reasons.

The exact power law behaviour in (16) implies that the $\rho(x)$ in (15)
cannot be normalized. Following \cite{l9} we introduce a simple cut-off
to get rid of the problem.
It could be also done in a more elegant way but for our purposes of rather
qualitative analysis it is enough to use the cut-off~:
\beq
\rho_{S}({\bf x})=L^{-\alpha} \mid{\bf x}\mid^{\alpha-3}
\Theta(L-\mid{\bf x}\mid)\alpha (4\pi)^{-1}
\eeq
In this case L can be interpreted as a size of the pion  production region.
The assumption of simultaneous particle production in (15) allows to consider
in (11) and respectively in (12), (13) only 3-dimensional Fourier transforms
of $\rho_{S}$. Using the relations for 3-dimensional Fourier transform~:
\beq
\rho_{S}({\bf k})=\int d^{3}x e^{-i{\bf kx}} \rho_{S}({\bf x})
\eeq
\beq
\rho_{S}({\bf k}=0)=\int d^{3}x \rho_{S}({\bf x})
\eeq
one obtains the distribution in momentum space~:
\beq
\rho_{S}({\bf k})=\alpha (L\mid{\bf k}\mid)^{-\alpha}
\int_{0}^{L\mid {\bf k}\mid } du \,u^{\alpha-2} \sin u
\eeq
One can observe {\sl the power law behaviour in (20) only for $L\mid {\bf k}
\mid \geq
1$. For $L\mid {\bf k} \mid \leq 1$ the singularity will be cut off,
and $\rho({\bf k})$
in (20) tends smoothly to 1 for $L\mid {\bf k}\mid \rightarrow 0$.
We notice also that $\rho_{S}({\bf k})\leq 1$ for any k.}

\subsection{Leading terms for multiplicities in the chaotic source current
ensemble}

Now we will discuss leading terms in (14) for pion multiplicities
defined in (11).
At first we consider the lowest order distributions (12)
and (13). After integrating the one-, two-, and three-pion distribution
from (15) in the region $(0,\delta)$ we get respectively~:
\beq
F_{1}(\delta)=<N>c_{11}(\delta)
\eeq
\beq
F_{2}(\delta)=<N^{2}>c_{21}(\delta)+<N(N-1)>c_{22}(\delta)
\eeq
\beq
F_{3}(\delta)=<N^{3}>c_{31}(\delta) + <N^{2}(N-1)>c_{32}(\delta) +
<N(N-1)(N-2)>c_{33}(\delta)
\eeq
where $c_{ij}$ are results of integrating $P_{1}({\bf k_{1}})$,
$P_{2}({\bf k_{1},k_{2}})$,
$P_{3}({\bf k_{1},k_{2},k_{3}})$ obtained from (11): they depend on $\delta$
{\sl but  they are independent of N.  It is also important to notice that
both terms
in (22) which survive  after phase averaging have the same rank of N.}
{}From (21) we obtain obviously $\alpha_{1}=0$ for $P_{1}({\bf k_{1} })$. From
the distribution $P_{2}({\bf k_{1},k_{2} })$ we will get the power
singularity
$\delta^{-2\alpha}$ only in $c_{22}(\delta)$. There are two singularities:
$c_{32}\sim \delta^{-2\alpha}$ and $c_{33}\sim \delta^{-3\alpha}$ in
$F_{3}$
of the same rank in N, so the singularity $\delta^{-2\alpha}$ will dominate
here. Hence the intermittency exponent is~:
\beq
f_{2}=f_{3}=2\alpha
\eeq
Now we will analyze the general form of multipion distribution
$P_{m}({\bf k_{1},\ldots,k_{m} })$
to get the dominating term  there.
At first let us consider terms which can survive in (11) after
phase averaging. The answer will be obtained from the analysis of the
expression~:
\beq
<e^{i(\varphi_{i_{1}}-\varphi_{i_{2}}) } e^{i(\varphi_{i_{3}}-\varphi_{i_{4}})}
\ldots e^{i(\varphi_{i_{2m-1}}-\varphi_{i_{2m}}) }>_{[\varphi_{i}]}
\eeq
where the average is taken over the set of phases $\varphi_{i}$.
The exponent product in  brackets should be equal to 1.
{\sl All the surviving terms have the same rank $N^{m}$ of N.}

So we are ready to consider leading terms in $P_{m}({\bf k_{1},\ldots,k_{m}})$
for any $m > 1$. Because all the terms have the same rank of N, and
$\rho_{S}({\bf k})$
shows the power law behaviour only for $L \delta \geq 1$ as mentioned in the
previous
section we conclude that the term with the smallest intermittency exponent
will
dominate and therefore~:
\beq
f_{m}= 2\alpha
\eeq
One can check this result holds on for both large and small N.
It means that {\sl intermittency exponents for factorial moments
will not change with the rank of multiplicity for the chaotic production}.
The result for scaled factorial moments will be the same.

The expression we have got above differs from intermittency exponents
obtained
in \cite{l11} for the coherent source ensemble. In \cite{l11} the
intermittency
exponents fulfilled the relation~:
\beq
f_{m}'=2m\alpha
\eeq
The calculations we have done to get $f'_{m}$ requested the assumption N to be
large enough. For small N the formulae were very complicated, and it was
difficult
to see how they could provide the simple power law behaviour
as observed in experiment. There were also no scaling in the scaled factorial
moments.

The result (26) can be actually derived under much less restrictive conditions.
To see this, let us first  formulate general conditions
that the function $\rho(x)$ must fulfill to be
considered as a probability distribution and to show the scaling behaviour~:
\beq
\rho_{S}({\bf x})\geq0
\eeq
\beq
\int\rho_{S}({\bf x})d^{3}x =1
\eeq
\beq
\rho_{S}({\bf k})\propto const(L\mid{\bf k}\mid)^{-\alpha} \,in
\,some \,interval
L\mid{\bf k}\mid \in (a,b)
\eeq
One can notice that conditions (28), (29) give the following inequality for
the 3-dimensional Fourier transform~:
\beq
\rho_{S}({\bf k})\leq \rho_{S}({\bf k=0})=1\, for\, any \,{\bf k}
\eeq
So we can generalize the results obtained for the collision sites
distribution
$\rho(x)$ defined in (15), (16). For any function $\rho(x)$ which fulfills
conditions
(28), (29), (30) intermittency exponents behave like (26).

\subsection{Correlation functions}

Analyzing the results from Ref.\ \cite{l11} for the  coherent source ensemble
one can observe that intermittency exponents (27) calculated from
multiplicities
are equal to intermittency  exponents obtained from
{\sl correlation functions}.
For the chaotic production we can make an interesting remark.
If the source number
N is fixed -- in our formalism it means that N is given by a "spike"
distribution $P(N)$ and $<N> \approx N$, the scaled correlation functions
look like~:
\beq
C_{2}({\bf k_{1},k_{2}})=\frac{N-1}{N}\mid\rho({\bf k_{1}-k_{2}})\mid^{2}
\eeq
\beq
C_{3}({\bf k_{1},k_{2},k_{3}})\,=\, \frac{(N-1)(N-2)}{N^{2}}\,[\,
\rho({\bf k_{1}-k_{2} })
\rho({\bf k_{3}-k_{1}})\rho({\bf k_{3}-k_{2}}) +c.c.]
\eeq
It seems to be possible to generalize easily the above results for
correlation functions
of any rank $m > 1$. Then intermittency exponents will grow with the rank m
following the rule~:
\beq
\nu_{m}= m\alpha
\eeq
identically with the behaviour of intermittency exponents calculated
in Ref.\ \cite{l9}.
And for multiplicities we get as usually intermittency exponents following
(26). There is no contradiction here. The term with the exponent
(34) appears also in the multiplicity but it will be small compared with
the term with the
smallest exponent (26). So correlation functions seem to be a better tool
to investigate the scaling properties of the pion production. In the
above case
the correlation function simply { \sl extracts} the term with the largest
exponent.


\section{Conclusions}

We have investigated the relation between intermittency and the scaling
behaviour
of the space-time distribution of collision sites in current ensemble
model. Our conclusions
can be summarized as follows~:

- there is a possibility of  intermittency in current ensemble model
provided that

I. source number N is not fixed and~:

 \begin{description}

 \item[(a)] one assumes a power law singularity in  the distribution
 of collision sites

 \item[(b)] the particle production is totally coherent

 \item[(c)] the number of coherent sources N is large enough.
 Then intermittency exponents grow linearly with the increasing rank of
 multiplicities
 following the Eq.\ (27). For small number of sources N intermittency is
 generally not observed. There are no well defined leading terms which can
 give
 intermittency exponents growing with the rank of moments/ multiplicities.
 In the scaled factorial moments/cumulants scaling does not appear.

 \end{description}

 II. source number N is fixed and~:

 \begin{description}

 \item[(a)] one assumes a power law singularity in the distribution of
 collision sites

 \item[(b)] the particle production is incoherent

 \end{description}

Then the intermittency exponents for scaled and not scaled factorial moments
do not change with
the rank of multiplicity. However, intermittency exponents for scaled
cumulants
follow the result obtained in Ref.\ \cite{l9} for one incoherent source,
i.e. the formula (34).

- the above results only partially agree with the experiment.
The intermittency exponents can grow with the rank of the
multiplicity/correlation
function, except the intermittency exponents for the factorial moments
in the case of incoherent production. In this case correlation functions
better detect scaling properties of the emission region than
multiparticle densities. However, the result for cumulants
$\nu_{3}=2 \nu_{2}$
obtained by the NA22 group cannot be confirmed.

\vspace{1.0cm}

I would like to thank Professor A. Bialas for many stimulating discussions,
suggestions and a continuous interest in this work. I am indebted to Professor
K. Fialkowski for his reading the manuscript and many helpful comments.

\end{document}